\documentclass[conference]{IEEEtran}
\IEEEoverridecommandlockouts
\usepackage[numbers,compress,comma]{natbib}
\usepackage{tcolorbox}
\usepackage{multirow}
\usepackage{tabulary} 
\usepackage{amsmath,amssymb,amsfonts}
\usepackage{algorithmic}
\usepackage{graphicx}
\usepackage{textcomp}
\usepackage{xcolor}
\usepackage{url}
\usepackage{flushend}
\usepackage{lipsum}

\usepackage{csquotes}

\SetBlockEnvironment{smallquote}
\newcommand{\aml}{AutoML}
\newcommand{\fullcite}[1]{\citeauthor{#1}~\cite{#1}}
\newcommand{\tildex}[1]{$_{\widetilde{~}}$#1}

\def\BibTeX{{\rm B\kern-.05em{\sc i\kern-.025em b}\kern-.08em
    T\kern-.1667em\lower.7ex\hbox{E}\kern-.125emX}}
\begin{document}

\title{Assessing the Use of AutoML for Data-Driven\\Software Engineering}

\author{\IEEEauthorblockN{Fabio Calefato, Luigi Quaranta, Filippo Lanubile}
\IEEEauthorblockA{\textit{University of Bari}\\
Bari, Italy \\
{ \{fabio.calefato, luigi.quaranta, filippo.lanubile\}@uniba.it}}
\vspace{-7mm}
\and
\IEEEauthorblockN{Marcos Kalinowski}
\IEEEauthorblockA{
\textit{PUC-Rio}\\
Rio de Janeiro, Brazil \\
kalinowski@inf.puc-rio.br}
\vspace{-7mm}
}

\IEEEpubid{978-1-6654-5223-6/23/\$31.00~\copyright~2023 IEEE ~~~~~~~~~~~~~~~~~~~~~~~~~~~~~~~~~~~~~~~~~~~~~~~~~~~~~~~~~~~~~~~~~~~~~~~~~~~~~~~~~~~~~~~~~~~~~~~~~~~~~~~~~~~~~~~~~~~~~~~~~~~~~~~~~~~}

\maketitle

\begin{abstract}
\textit{Background}. 
Due to the widespread adoption of Artificial Intelligence (AI) and Machine Learning (ML) for building software applications, companies are struggling to recruit employees with a deep understanding of such technologies.
In this scenario, \aml\ is soaring as a promising solution to fill the AI/ML skills gap since it promises to automate  the building of end-to-end AI/ML
pipelines that would normally be engineered by specialized team members.
\textit{Aims}. Despite the growing interest and high expectations, there is a dearth of information about the extent to which \aml\ is currently adopted by teams developing AI/ML-enabled systems and how it is perceived by practitioners and researchers. 
\textit{Method}. To fill these gaps, in this paper, we present a mixed-method study comprising a benchmark of 12 end-to-end \aml\ tools on two SE datasets and a user survey with follow-up interviews to further our understanding of \aml\ adoption and perception.
\textit{Results}. We found that \aml\ solutions can generate models that outperform those trained and optimized by researchers to perform classification tasks in the SE domain.
Also, our findings show that the currently available \aml\ solutions do not live up to their names as they do not equally support automation across the stages of the ML development workflow and for all the team members. 
\textit{Conclusions}. We derive insights to inform the  SE research community on how \aml\ can facilitate their activities and tool builders on how to design the next generation of \aml\ technologies.
\end{abstract}


\section{Introduction}\label{sec:introduction}
The recent advancements in Artificial Intelligence (AI) and Machine Learning (ML) have led to their widespread industrial adoption in a variety of domains, such as automotive, business, and healthcare~\cite{Jordan2015,Luckow2018}.
Building well-performing AI-augmented applications requires more than just highly specialized human experts with a deep knowledge of AI/ML-related technologies.
Working in specialized teams, data engineers, domain experts, statisticians, and software engineers help data scientists
develop AI/ML-enabled systems by building end-to-end pipelines comprising a variety of stages, from data preprocessing and feature engineering to model integration, monitoring, and fine-tuning.
However, companies are currently struggling to recruit AI/ML experts~\cite{Markow2017,Bhaskar2021,Zhang2021}.

Given this severe AI/ML skills shortage, \textit{automated machine learning} (\aml)~\cite{Hutter2019} has gained momentum as it aims to fill this gap by automatizing the building of end-to-end AI/ML pipelines that would normally be engineered by specialized teams. 
\aml\ is in fact appealing to both companies lacking in-house expertise as well as those that already have specialists who thus benefit from automation when executing complex, time-consuming activities such as data exploration, feature engineering, and hyperparameter optimization.
Lately, \aml\ has also been employed in Software Engineering (SE) research; 
\fullcite{Tanaka2019} experimented with AutoML to develop software defect prediction models, achieving state-of-the-art performances.

Notwithstanding the growing interest and high expectations, we have little knowledge regarding the extent to which \aml\ is currently adopted by teams developing AI/ML-enabled systems.
\fullcite{Vanschoren2014} found that \aml\ accounted for less than 2\% of the workflows uploaded to OpenML,\footnote{\url{https://openml.org}} an online platform collecting public ML datasets and benchmark results.
\fullcite{vanderBlom2021} conducted a survey with members of AI/ML teams, which revealed that about 20-30\% of the respondents had never used \aml.

Moreover, we have a limited grasp of how \aml\ is perceived by practitioners. 
In particular, \fullcite{Wang2019} have explored the perceptions of adopting \aml, focusing on the perspective of data scientists and how they use it for building models and fine-tuning hyperparameters.
Nonetheless, \aml\ tools aim to cover the AI/ML workflow end to end, and little is still known about the perception and expectations of other team members.
Although the boundaries are often blurred, a mature AI/ML team consists of several roles other than those pertaining to data scientists such as \textit{data engineers}, who focus on setting up data preparation pipelines, and \textit{ML engineers} (often referred to also as MLOps), who focus on engineering activities like model integration, deployment, and monitoring~\cite{Sato2019,Menzies2019}.
In~\cite{Wang2023}, \citeauthor{Wang2023} analyzed several Stack Overflow questions and GitHub discussions to define a taxonomy of 26 developer-specific challenges when using \aml.
In addition, \fullcite{Majidi2022} sampled \tildex{22k} GitHub projects adopting \aml\ tools and found that they are mostly used for model training and evaluation.

Finally, 
there are several published benchmarks comparing the performance of \aml\ systems (e.g., \cite{Gijsbers2019,Guyon2019,Zoller2021}); for example, research has reported that \aml\ does not outperform conventional forecasting strategies used in time series analysis~\cite{Paldino2021}.
However, with the exception of \fullcite{Tanaka2019}, none of these benchmarks have leveraged datasets from the SE domain so far, notwithstanding the several potential applications of \aml\ in this field.

To bridge these gaps, in this paper, we present a mixed-method study comprising (i) a benchmark of 12 end-to-end \aml\ tools on two software engineering datasets, to collect initial evidence about the performance of state-of-the-art \aml\ solutions when applied to SE-specific tasks -- namely sentiment analysis from technical text -- and (ii) a user survey with follow-up interviews with software engineers working on AI/ML projects, to further our understanding of their perception and extent of \aml\ adoption.

Our contributions are appealing to both industry and the SE research community as, respectively, we (i) show what stages of a typical ML workflow are more/less suitable for automation according to SE practitioners and (ii) provide evidence that \aml\ solutions may generate models that outperform those trained and optimized by researchers to perform classification tasks -- such as sentiment analysis and emotion recognition in the SE domain -- and at the same time reach performance close to complex and time-consuming solutions based on fine-tuning pre-trained transformers (e.g., BERT~\cite{Devlin2018}).

The remainder of this paper is organized as follows. 
In Sect.~\ref{sec:background}, we review background literature.
In Sect.~\ref{sec:study-design}, we illustrate the design of the mixed-method study.
The results of the quantitative and qualitative analyses are reported and discussed in Sect.~\ref{sec:results} and \ref{sec:discussion}, respectively.
Finally, we draw conclusions in Sect.~\ref{sec:conclusions}.

\section{Background}\label{sec:background}

\subsection{Machine Learning Development Workflow}\label{sec:workflow}

\begin{figure}
    \centering
    \includegraphics[width=9cm]{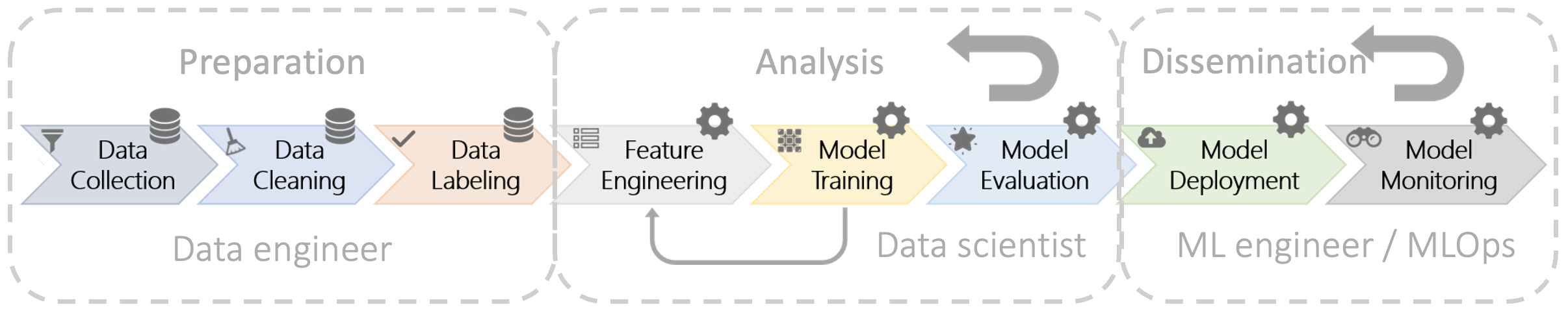}
    \vspace{-7mm}
    \caption{A typical ML workflow with activities and roles (adapted from \fullcite{Amershi2019}).}
    \vspace{-3mm}
    \label{fig:ml-workflow}
\end{figure}

Fig.~\ref{fig:ml-workflow}  presents a typical machine-learning workflow with activities and responsibilities.
Compared to a generic, say agile, development workflow of a traditional software-engineered system two main differences arise. 
First, with AI/ML-enabled systems, the final output of the development workflow is not the code itself, but rather the models' outcome (e.g., the predictions made, the informed business decisions taken)~\cite{Epperson2022}.
The second difference pertains to the amount of experimentation needed to converge to a satisfying model performance for the problem at hand; in fact, albeit the workflow may appear linear, much like agile software processes, a machine learning workflow contains several feedback loops due to the frequent iterations involving model selection, hyper-parameters tuning, and dataset refinement~\cite{Amershi2019}.

\textbf{Preparation}. The data preparation stage is typically the responsibility of data engineers.
If no data is already available, the first step in the preparation process is \textit{data collection}, where data engineers retrieve raw pieces of information from multiple data sources and then integrate them into a dataset.
Once acquired, the next step to perform with the dataset is \textit{data cleaning}, which involves removing inaccurate or noisy records, a common activity in all forms of data science. 
The typical preprocessing steps regard the imputation of missing values as well as removing duplicates and inconsistent values. 
The last step in data preparation is \textit{data labeling}, the time-intensive task of assigning ground truth labels to each record in a dataset since most of the supervised learning techniques require `gold' labels to be able to build a model. 
This step refers to both `assisted data labeling' and `manual data labeling,' depending on whether ML is used 
to complete the annotation and save time and budget, as compared to only using human annotators (e.g., data engineers themselves, domain experts, and crowd workers) to label the entire dataset.

\textbf{Analysis}. This stage is the core of a machine learning development workflow, which typically pertains to data scientists who perform several tasks to eventually build and release a validated model.
The first step in the analysis process is \textit{feature engineering}, which refers to all those activities performed to extract and select from the dataset the most informative features for building ML models.
The typical feature engineering tasks include normalizing and discretizing  numeric features, creating  one-hot encoding and embeddings for categorical features, basic processing such as tokenization for text features, and extracting date- and time-related features from timestamped columns.
For deep learning models (e.g. convolutional neural networks), feature engineering is less explicit and often blended with the next stage.
During \textit{model training} the chosen models are trained on the annotated dataset, using the selected features.
Once built, models are evaluated in the \textit{model evaluation} using  predefined performance metrics (e.g., log-loss, accuracy, AUC).
This is a critical step that typically involves extensive human evaluation, with data scientists assessing the output models to ensure they meet the required performance criteria.

\textbf{Dissemination}. The final stage of the machine learning development workflow typically pertains to ML engineers and software developers.
\textit{Model deployment} is the step where the evaluated machine learning model is integrated into targeted production environments to make practical predictions or business decisions based on new data.
During \textit{model monitoring}, engineers continuously monitor the deployed models for possible errors and decrease in performance during real-world execution (e.g., model drift).

\subsection{AutoML}\label{sec:aml}
\aml\ is the portmanteau of \textit{Automated} and \textit{Machine Learning}. 
The term is generally used to refer to solutions aiming at democratizing ML to non-experts by easing the difficulty of developing ML models and, at the same time, drastically increasing the productivity of experts.

\aml\ has been gaining a lot of momentum in recent years, especially  after the release of commercial solutions from large companies such as Amazon, Microsoft, and IBM.
Despite the recent interest, \aml\ is not entirely a new trend because, since the 1990s, many  solutions have been proposed for automatic Combined Algorithm Selection and Hyperparameter (CASH) optimization~\cite{Zoller2021}.
With respect to the workflow depicted in Fig.~\ref{fig:ml-workflow}, \aml\ solutions typically focus on automating the `core' tasks of the Analysis stage (i.e., feature engineering, model training, and evaluation), with less attention devoted to preparation and dissemination (we further discuss this in Sect.~\ref{sec:tools}, where we describe the tools selected for our analysis).

Research in the area of \aml\ has also gained significant traction leading to many improvements in terms of both performance and reduced execution runtime~\cite{Hutter2019}.
Researchers have published several benchmarks that compare the performance of models trained on datasets from various domains using \aml\ solutions. 
The reported findings are varied.
\fullcite{Paldino2021} found that \aml\ does not outperform conventional forecasting strategies used in time series analysis.
\fullcite{Hanussek2020} used 4 \aml\ tools to perform regression and classification tasks on 12 different popular datasets from OpenML and found that the automated frameworks perform better than or equal to the machine learning community in 7 out of 12 cases.
\fullcite{Ferreira2021} report similar findings in a study comparing the performance of 8 open-source \aml\ tools on 12 OpenML datasets. 
In a similar study conducted on a broader set of 57 classification datasets and 30 regression datasets retrieved from OpenML, \fullcite{Balaji2018} report high variances in the results.
\fullcite{Gijsbers2019} built a framework to benchmark \aml\ solutions and conducted a comparison of 4 systems across 39 datasets against a Random Forest classifier used as baseline; with enough time budged (4h), none of the \aml\ solutions outperformed the Random Forest classifier.
\fullcite{Guyon2019} found a performance gap between human-tweaked and \aml-generated model pipelines retrieved from various editions of the `\aml\ Challenge.'
Finally, in the SE domain, \fullcite{Tanaka2019} studied the effectiveness of \aml\ in predicting the number of defects in software modules; using auto-sklearn, they achieved a prediction performance similar to those reported in prior studies.

\vspace{-0.4mm}
\subsection{XAI}

AI/ML techniques are in general not easy to understand, even for experts.
This has led to a growing interest in the topic of XAI~\cite{Gunning2019xai}, i.e., techniques focusing on \textit{interpretability} (i.e., the understandability of the intuition behind a model's output) and \textit{explainability} (i.e., the understandability of a model's internal logic and mechanics)~\cite{Linardatos2021}.
Model explanations may apply to a single prediction (local) or to a model's entire behavior (global).
Furthermore, interpretability can be \textit{direct}, to indicate the use of self-explainable models whose internal logic is intrinsically transparent and therefore generally understandable (e.g.,  decision trees), or \textit{post-hoc}, which involves the use of an auxiliary method to explain a complex, black-box model's behavior (e.g.,  feature importance scores, heatmaps, and natural language descriptions of its inner working and decision logic).
Interpretability features may also apply to any type of algorithm (agnostic) or only to one family (model-specific).
Finally, fairness indicates the presence of features to detect and deal with biases (e.g., ethnicity, gender, disability status) in models and training data. 

\section{Study Design}\label{sec:study-design}
In this work, we investigate the use of \aml\ tools for SE.
In particular, we ask the two following research questions:
\begin{itemize}
     \item \noindent \textbf{RQ1} -- \textit{How do \aml\ tools perform on SE tasks?}
    \item \noindent \textbf{RQ2} -- \textit{How do software engineers currently use \aml? How does it fit their workflow? What are their perceptions?}
\end{itemize}

\begin{figure}[t]
    \centering
    \vspace{-1mm}
    \includegraphics[width=5.5cm]{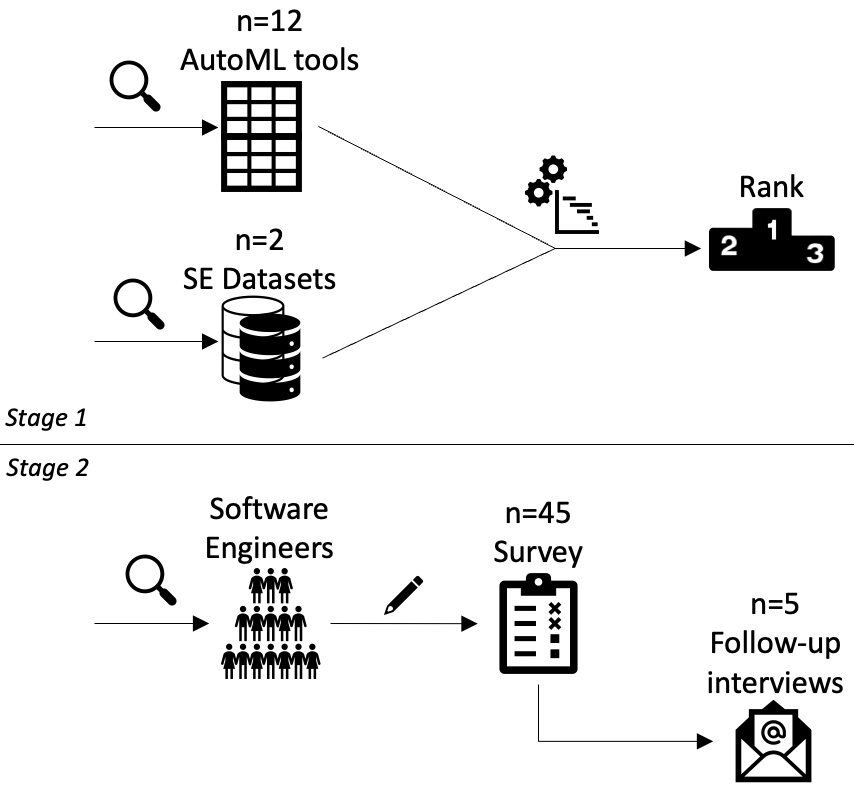}
    \vspace{-4mm}
    \caption{The two experimental stages executed in the study.}
    \vspace{-4mm}
    \label{fig:study}
\end{figure}

To answer the questions, we follow a mixed-methods approach
as depicted in the two-stage research framework in Fig.~\ref{fig:study}.
In \textit{stage 1}, to answer RQ1 we perform a quantitative assessment of \aml\ tools performance. 
Accordingly, we first select \textit{12} \aml\ solutions, both open-source (4) and commercial (8).
The list of selected tools is reported in Table~\ref{tab:tools} (see Sect.~\ref{sec:tools}).
Then, we choose a text-classification task -- i.e., sentiment analysis from technical text -- which has recently gained considerable attraction in the SE research community;  
contextually, we retrieve two datasets totaling over 10k documents written by developers (Sect.~\ref{sec:datasets}). 
Finally, we apply the selected tools to the datasets and compare their performance using the same metrics adopted in previous benchmark studies that developed SE-specific tools for the tasks (Sect.~\ref{sec:metrics}).
In \textit{stage 2}, 
to answer RQ2 we perform a qualitative analysis aimed to garner insights about the current extent of adoption of \aml\ at the various stages of the AI/ML workflow and the potential different perceptions of software engineers about using \aml\ solutions. 
Accordingly, we carried out an online survey with 45 participants; we also conducted follow-up interviews with 5 participants to deepen the relevant themes that emerged from the survey analysis (Sect.~\ref{sec:survey}).

\subsection{AutoML Tools}\label{sec:tools}

Table~\ref{tab:tools} lists the main characteristics of the tools sampled for conducting the \aml\ benchmark.
The process of sampling the tools was conducted in two steps.

First, we selected commercial solutions.
As a seed, we started by including the \aml\ solutions provided by the companies identified as leaders in the \textit{Gartner's Magic Quadrant For Data Science And ML Platforms}, namely Amazon, Google, IBM, and Microsoft.
Then, we performed a web search on Google using the keywords `\textit{automl}' and `\textit{autoai}' and selected the first four commercial products found, thus doubling the number of those already included (8 overall).

Second, to identify open-source solutions, we explored the GitHub topics pages for the keywords `\textit{automl}' and `\textit{autoai};' overall, we identified 831 repositories.
Because OSS solutions are generally libraries that require at least some coding, we opportunistically chose to retain only tools written in  Python (395).
Furthermore, given the popularity of the topic, we selected the  repositories having 200+ stars (99), to ensure removing small or personal projects. 
Also, we removed: (i) `stale' projects, retaining only the repositories updated in the last twelve months (79), deemed a sufficient amount of time, considering the hype around the topic; (ii) false positives, i.e., repositories not actually containing \aml\ tools (38); (iii) repositories with README files missing or not in English (34);  (iv) libraries supporting only Computer Vision tasks or a specific workflow activity, such as hyperparameter optimization (HPO), neural architectural search (NAS), and model compression (28); (v) smaller-scale solutions from vendors of larger commercial solutions already selected in the previous step (25); (vi) repositories containing resources for educational purposes or competitions (23).
Finally, we manually inspected the existing documentation of the remaining tools and retained only those supporting both text features (NLP) and providing straightforward examples of how to automate a text analysis task 
(see the experimental datasets and task descriptions in Sect.~\ref{sec:datasets}).
At the end of this step, we selected four open-source \aml\ tools (marked with a $^*$ in Table~\ref{tab:tools}).

Next, we briefly comment on the characteristics of the selected twelve tools, grouped by type.


\textbf{Open-source solutions}. We observe that open-source solutions can only be run on-premise and can only be used as APIs (i.e., requiring code development) or command-line applications (AutoGoal and Ludwig). All solutions support image, tabular, and textual data types. 
With respect to the stages and activities highlighted in the typical ML workflow in Fig.~\ref{fig:ml-workflow}, all these solutions fully support the analysis stage as the core activities of ML-enabled systems development.
Regarding preparation, they support data cleaning but provide no support for ML-assisted data labeling.
Also, with respect to the dissemination stage, none of the sampled solutions support model deployment and monitoring, with the sole exception of Ludwig, which allows model pipelines to be served by spawning a REST API with one command.
Finally, none of the open-source solutions currently supports explainability.

\textbf{Commercial solutions}. The landscape of commercial \aml\ solutions is different than the previous one. 
First, these solutions run on cloud infrastructure with the sole exception of RapidMiner Studio, which is also the only desktop solution; all the other platforms are web applications also offering programmatic access via APIs.
The model analysis stage is also well supported by commercial solutions. 
Compared to OSS solutions, they better support the preparation stage, by also providing support for assisted data labeling, and dissemination, as they all allow one-click model deployment and monitoring.

In addition, with the exception of RapidMiner Studio, all the benchmarked commercial \aml\ platforms  support XAI, albeit to a slightly different extent.
We found that, in general, these solutions ensure fairness by providing features to spot model/data biases and implement both model-dependent and independent (agnostic) features.
In terms of explanation scope, they offer both local and global explanations.
Regarding the support to interpretability, self-explaining models are generally supported thus, with the exception of BigML, these tools focus on providing \textit{post-hoc} explanation.


Wrapping up, we note that, as compared to open-source solutions, commercial \aml\ platforms offer code-free and more end-to-end solutions -- as they cover the dissemination stage, offering model deployment and monitoring automation -- and also support XAI to various extents.

\begin{table*}[t]
\centering
\vspace{-6mm}
\caption{A breakdown of the characteristics of \aml\ solutions benchmarked in this study 
(*=open source, ?=feature absence/presence inferred upon code inspection, not from documentation).}
\vspace{-3mm}
\label{tab:tools}
\begin{tabular}{lccccccccccc}
\hline
\multirow{2}{*}{\textbf{Tool}}                                          & \multirow{2}{*}{\textbf{\begin{tabular}[c]{@{}c@{}}Infra\\ struct.\end{tabular}}} & \multirow{2}{*}{\textbf{Solution}}                 & \multirow{2}{*}{\textbf{\begin{tabular}[c]{@{}c@{}}Data \\ types\end{tabular}}} & \multicolumn{2}{c}{\textbf{Preparation}}                                                                                        & \multicolumn{3}{c}{\textbf{Analysis}}                                                                                                                                                                  & \multicolumn{2}{c}{\textbf{Dissemination}}                                                                                            & \multirow{2}{*}{\textbf{XAI}} \\ \cline{5-11}
                                                                        &                                                                                   &                                                    &                                                                                 & \textbf{\begin{tabular}[c]{@{}c@{}}Data\\ clean.\end{tabular}} & \textbf{\begin{tabular}[c]{@{}c@{}}Data\\ label.\end{tabular}} & \textbf{\begin{tabular}[c]{@{}c@{}}Feature\\ engin.\end{tabular}} & \textbf{\begin{tabular}[c]{@{}c@{}}Model\\ traninig\end{tabular}} & \textbf{\begin{tabular}[c]{@{}c@{}}Model\\ eval.\end{tabular}} & \textbf{\begin{tabular}[c]{@{}c@{}}Model\\ deploym.\end{tabular}} & \textbf{\begin{tabular}[c]{@{}c@{}}Model\\ monitor.\end{tabular}} &                               \\ \hline
AutoGluon*                                                              & On prem.                                                                          & API                                                & img, tab, txt                                                                   & Y                                                              & N                                                              & Y                                                                 & Y                                                                 & Y                                                              & N                                                                 & N                                                                 & N                             \\ \hline
AutoGoal*                                                               & On prem.                                                                          & \begin{tabular}[c]{@{}c@{}}API,\\ CLI\end{tabular} & img, tab, txt                                                                   & N?                                                             & N                                                              & Y?                                                                & Y                                                                 & Y                                                              & N                                                                 & N                                                                 & N                             \\ \hline
AutoKeras*                                                              & On prem.                                                                          & API                                                & img, tab, txt                                                                   & Y?                                                             & N                                                              & Y?                                                                & Y                                                                 & Y                                                              & N                                                                 & N                                                                 & N                             \\ \hline
\begin{tabular}[c]{@{}l@{}}Amazon \\ SageMaker\\ AutoPilot\end{tabular} & Cloud                                                                             & \begin{tabular}[c]{@{}c@{}}API,\\ Web\end{tabular} & img, tab, txt                                                                   & Y                                                              & Y                                                              & Y                                                                 & Y                                                                 & Y                                                              & Y                                                                 & Y                                                                 & Y                             \\ \hline
BigML                                                                   & Cloud                                                                             & \begin{tabular}[c]{@{}c@{}}API,\\ Web\end{tabular} & img, tab, txt                                                                   & Y                                                              & N                                                              & Y                                                                 & Y                                                                 & Y                                                              & Y                                                                 & Y                                                                 & Y                             \\ \hline
\begin{tabular}[c]{@{}l@{}}DataRobot \\ AI Cloud\end{tabular}           & Cloud                                                                             & \begin{tabular}[c]{@{}c@{}}API,\\ Web\end{tabular} & img, tab, txt                                                                   & Y                                                              & Y                                                              & Y                                                                 & Y                                                                 & Y                                                              & Y                                                                 & Y                                                                 & Y                             \\ \hline
\begin{tabular}[c]{@{}l@{}}Google \\ Vertex AI\end{tabular}             & Cloud                                                                             & \begin{tabular}[c]{@{}c@{}}API,\\ Web\end{tabular} & \begin{tabular}[c]{@{}c@{}}img, tab,\\ txt, vid\end{tabular}                    & Y                                                              & Y                                                              & Y                                                                 & Y                                                                 & Y                                                              & Y                                                                 & Y                                                                 & Y                             \\ \hline
\begin{tabular}[c]{@{}l@{}}H20 \\ Driverless AI\end{tabular}            & \begin{tabular}[c]{@{}c@{}}Cloud,\\ On prem.\end{tabular}                         & \begin{tabular}[c]{@{}c@{}}API,\\ Web\end{tabular} & img, tab, txt                                                                   & Y                                                              & Y                                                              & Y                                                                 & Y                                                                 & Y                                                              & Y                                                                 & Y                                                                 & Y                             \\ \hline
\begin{tabular}[c]{@{}l@{}}IBM \\ Watson\\ AutoAI\end{tabular}          & Cloud                                                                             & \begin{tabular}[c]{@{}c@{}}API,\\ Web\end{tabular} & img, tab, txt                                                                   & Y                                                              & N                                                              & Y                                                                 & Y                                                                 & Y                                                              & Y                                                                 & Y                                                                 & Y                             \\ \hline
\begin{tabular}[c]{@{}l@{}}Ludwig AI \\ AutoML*\end{tabular}            & On prem.                                                                          & \begin{tabular}[c]{@{}c@{}}API,\\ CLI\end{tabular} & img, tab, txt                                                                   & Y?                                                             & N                                                              & Y                                                                 & Y                                                                 & Y                                                              & Y                                                                 & N                                                                 & N                             \\ \hline
\begin{tabular}[c]{@{}l@{}}MS Azure \\ AutoML\end{tabular}              & Cloud                                                                             & \begin{tabular}[c]{@{}c@{}}API,\\ Web\end{tabular} & img, tab, txt                                                                   & Y                                                              & Y                                                              & Y                                                                 & Y                                                                 & Y                                                              & Y                                                                 & Y                                                                 & Y                             \\ \hline
Rapid Miner Studio             & On prem.                                                                          & Desktop                                            & img, tab, txt                                                                   & Y                                                              & N                                                              & Y                                                                 & Y                                                                 & Y                                                              & Y                                                                 & Y                                                                 & N                             \\ \hline
\end{tabular}
\vspace{-3mm}
\end{table*}


\subsection{Datasets}\label{sec:datasets}

We selected two text-classification datasets from the SE domain.
The first dataset is retrieved from  \fullcite{Novielli2020} and contains 7,122 sentences from GitHub pull-request and commit comments
manually annotated with sentiment polarity labels based on Shaver et al.'s framework~\cite{Shaver1987} (28.3\% positive, 29.3\% negative, and 42.4\% neutral).
The second dataset is retrieved from \fullcite{Ortu2016} and includes about 6,000 issue comments and sentences authored by software developers of popular open-source software projects, such as Apache and Spring. 
This Jira dataset is manually annotated with six discrete emotion labels also from the Shaver et al.'s framework~\cite{Shaver1987} (i.e., love, joy, anger, fear, surprise, and sadness), whereas this study focuses on emotion polarity (i.e., the positive, negative, or neutral valence conveyed by texts).
Therefore, positive emotions, i.e., `love' and `joy,' are translated into a positive polarity label. 
Similarly, `sadness,' `anger,' and `fear' are mapped to the negative polarity class. 
Instead, the cases labeled as `surprise' are discarded because this emotion label could be either considered positive or negative, depending on the expectations of the author of a text.
Unlike the GitHub dataset, the Jira dataset is not well-balanced (31.3\% positive and 68.7\% negative).


\subsection{Metrics}\label{sec:metrics}

To evaluate the \aml\ tools' performance, we rely on the same metrics of Precision (P), Recall (R), and F1-measure (F) used in previous works, thus enabling a performance comparison.
We also report macro- and micro-averaged metrics to show overall classification performance scores across the classes.
Macro-averaging averages the metrics equally over the classes, whereas micro-averaging calculates the scores over the data points in all classes, thus it tends to be influenced by the performance of the majority class~\cite{Sebastiani2002}.

Because tools can be configured to generate models that maximize the classification accuracy with respect to a specific metric, to enable fair comparison we choose to optimize with respect to F1-measure.

Regarding the sentiment polarity dataset, the same metrics were used by \fullcite{Calefato2018} to assess the performance of Senti4SD, a SE-specific classifier, for each of the three classes (positive, negative, and neutral). 
Regarding the emotion dataset, instead, \fullcite{Ortu2016} did not provide an assessment of classification performance along with the Jira gold standard; however, a classification performance in terms of Precision, Recall, and F-measure is reported in the benchmark study by \fullcite{Novielli2020}.

For the sake of completion, we also compare our results to those reported in \fullcite{Zhang2020}, who show that fine-tuned pre-trained transformer-based models (e.g., BERT) outperform the existing SE-specific sentiment analysis tools.

 
Finally, to ensure a fair comparison among the tools, we manually changed their configuration, thus creating a shared experimental setting  resembling that of prior work by: (i) using a common optimization metric during the model analysis stage---we opted for maximizing F1-score; (ii) setting the same train/test data split---specifically, 70\% training, 20\% validation, and 10\% test.
Because it was not always possible to set a time budget, we ignore the time performance in our benchmark as execution times are not comparable.

\subsection{Survey \& Interviews}\label{sec:survey}

\textbf{Recruitment}. To recruit survey participants, we sought ML and software engineers with real-world experience on \aml. 
We recruited participants through public announcements posted on Twitter, relevant subreddits, and LinkedIn groups as well as through personal contact networks.
No monetary compensation was offered.

\textbf{Survey}. To create and host the web survey we used Google Forms.
Overall, the survey contained 43 questions structured into four parts (see the supplemental online material).\footnote{\url{https://doi.org/10.6084/m9.figshare.22640248}} 
The first two parts are intended to collect the participants' demographic.
We asked respondents to report their overall working experience,  experience with AI/ML-based systems development, their job role as well as typical team size, company characteristics, typical project type, and development workflow; participants having no experience whatsoever with AI/ML projects were redirected to  a disqualification page.
Furthermore, we screened participants based on their familiarity and experience with \aml\ and collected information about their general attitudes toward using \aml\ tools. 
The third part was conditionally filled out by respondents who reported having some familiarity and experience with \aml; it collected feedback on how \aml\ currently fits their development workflows, in general, and their specific task(s), in particular, as well as points of friction and challenges, if any. 
The fourth part was filled out by all survey respondents as it aimed to collect insights about how participants would want to use it, what workflow stages and steps they wish \aml\ tools supported better, and any features missing.
The survey ended with an open-ended question to collect any other feedback from the participants and their consent to be contacted for a follow-up interview.
Finally, two of the authors engaged in a process of inductive coding to extract and iteratively synthesize through discussion the common themes that arose from both the analysis of the open-ended questions.

\textbf{Follow-up interviews}. We conducted follow-up interviews with survey respondents who gave consent; they agreed to further elaborate on their answers and reflect on the themes that emerged from the survey analysis over email or chat.
After completing the interviews, two of the authors analyzed the written answers to identify and extract excerpts that were deemed to expand further the relevant themes identified from the analysis of the open-ended answers.

\begin{table*}[!ht]
\vspace{-6mm}
\caption{Results of the benchmark on the two datasets. The best results in terms of micro- and macro-averages are shown in bold (note that in multi-class classifications where each data point is assigned to exactly one class, the micro-avg of Precision, Recall, and F-measure hold the same value).}
\vspace{-2mm}
\label{tab:results}
\begin{tabular}{cc|ccc|ccc|ccc|ccc|c}
\hline
\multirow{2}{*}{\textbf{Tool}}                                                           & \multirow{2}{*}{\textbf{\begin{tabular}[c]{@{}c@{}}Dataset\\ (algorithm)\end{tabular}}} & \multicolumn{3}{c|}{\textbf{Positive}} & \multicolumn{3}{c|}{\textbf{Neutral}} & \multicolumn{3}{c|}{\textbf{Negative}} & \multicolumn{3}{c|}{\textbf{Macro-avg}} & \multirow{2}{*}{\textbf{Micro-avg}} \\ \cline{3-14}
                                                                                         &                                                                                         & \textbf{P}  & \textbf{R}  & \textbf{F} & \textbf{P}  & \textbf{R} & \textbf{F} & \textbf{P}  & \textbf{R}  & \textbf{F} & \textbf{P}  & \textbf{R}  & \textbf{F}  &                                     \\ \hline
\multirow{2}{*}{AutoGluon}                                                               & \begin{tabular}[c]{@{}c@{}}GitHub\\ (Weight. Ensem.)\end{tabular}                       & 0.73        & 0.73        & 0.73       & 0.78        & 0.85       & 0.81       & 0.74        & 0.63        & 0.68       & 0.75        & 0.74        & 0.74        & 0.75                                \\
                                                                                         & \begin{tabular}[c]{@{}c@{}}Jira\\ (Weight. Ensem.)\end{tabular}                         & 0.77        & 0.85        & 0.81       & 0.90        & 0.90       & 0.90       & 0.76        & 0.67        & 0.71       & 0.81        & 0.81        & 0.81        & 0.86                                \\ \cline{2-15} 
\multirow{2}{*}{AutoGoal}                                                                & \begin{tabular}[c]{@{}c@{}}GitHub\\ (Perceptron)\end{tabular}                           & 0.84        & 0.79        & 0.81       & 0.82        & 0.82       & 0.82       & 0.77        & 0.82        & 0.80       & 0.81        & 0.81        & 0.81        & 0.81                                \\
                                                                                         & \begin{tabular}[c]{@{}c@{}}Jira\\ (LinearSVC)\end{tabular}                              & 0.54        & 0.71        & 0.61       & 0.84        & 0.79       & 0.66       & 0.72        & 0.61        & 0.66       & 0.70        & 0.70        & 0.70        & 0.75                                \\ \cline{2-15} 
\multirow{2}{*}{AutoKeras}                                                               & \begin{tabular}[c]{@{}c@{}}GitHub\\ (Neural Net.)\end{tabular}                          & 0.91        & 0.95        & 0.93       & 0.93        & 0.88       & 0.90       & 0.89        & 0.91        & 0.90       & 0.91        & 0.91        & 0.91        & 0.91                                \\
                                                                                         & \begin{tabular}[c]{@{}c@{}}Jira\\ (Neural Net.)\end{tabular}                            & 0.80        & 0.86        & 0.83       & 0.92        & 0.89       & 0.91       & 0.75        & 0.80        & 0.77       & 0.82        & \textbf{0.85}        & \textbf{0.83}        & \textbf{0.87}                                \\ \cline{2-15} 
\multirow{2}{*}{\begin{tabular}[c]{@{}c@{}}Amazon \\ SageMaker\\ Autopilot\end{tabular}} & \begin{tabular}[c]{@{}c@{}}GitHub\\ (XGB)\end{tabular}                                  & 0.74        & 0.63        & 0.68       & 0.70        & 0.84       & 0.77       & 0.73        & 0.63        & 0.67       & 0.73        & 0.70        & 0.71        & 0,72                                \\
                                                                                         & \begin{tabular}[c]{@{}c@{}}Jira\\ (XGB)\end{tabular}                                    & 0.79        & 0.80        & 0.79       & 0.88        & 0.92       & 0.90       & 0.88        & 0.66        & 0.75       & \textbf{0.85}        & 0.80        & 0.82        & 0.86                                \\ \cline{2-15} 
\multirow{2}{*}{BigML}                                                                   & \begin{tabular}[c]{@{}c@{}}GitHub\\ (Decision Forest)\end{tabular}                      & 0.78        & 0.59        & 0.67       & 0.65        & 0.85       & 0.74       & 0.75        & 0.62        & 0.68       & 0.73        & 0.69        & 0.70        & 0.70                                \\
                                                                                         & \begin{tabular}[c]{@{}c@{}}Jira\\ (Decision Forest)\end{tabular}                        & 0.77        & 0.83        & 0.80       & 0.90        & 0.91       & 0.90       & 0.88        & 0.68        & 0.77       & \textbf{0.85}        & 0.81        & 0.82        & \textbf{0.87}                                \\ \cline{2-15} 
\multirow{2}{*}{\begin{tabular}[c]{@{}c@{}}DataRobot\\ AI Cloud\end{tabular}}            & \begin{tabular}[c]{@{}c@{}}GitHub\\ (CNN)\end{tabular}                                  & 0.95        & 0.90        & 0.92       & 0.90        & 0.95       & 0.93       & 0.91        & 0.88        & 0.90       & \textbf{0.92}        & 0.91        & \textbf{0.92}        & \textbf{0.92}                                \\
                                                                                         & \begin{tabular}[c]{@{}c@{}}Jira\\ (CNN)\end{tabular}                                    & 0.78        & 0.79        & 0.78       & 0.90        & 0.92       & 0.91       & 0.82        & 0.70        & 0.76       & 0.83        & 0.80        & 0.82        & \textbf{0.87}                                \\ \cline{2-15} 
\multirow{2}{*}{\begin{tabular}[c]{@{}c@{}}Google \\ Vertex AI\end{tabular}}             & \begin{tabular}[c]{@{}c@{}}GitHub\\ (?)\end{tabular}                                    & 0.93        & 0.92        & 0.93       & 0.88        & 0.92       & 0.90       & 0.88        & 0.83        & 0.86       & 0.90        & 0.89        & 0.89        & 0.89                                \\
                                                                                         & \begin{tabular}[c]{@{}c@{}}Jira\\ (?)\end{tabular}                                      & 0.77        & 0.91        & 0.83       & 0.90        & 0.91       & 0.91       & 0.86        & 0.60        & 0.70       & 0.84        & 0.80        & 0.82        & \textbf{0.87}                                \\ \cline{2-15} 
\multirow{2}{*}{\begin{tabular}[c]{@{}c@{}}H2O \\ Driverless AI\end{tabular}}            & \begin{tabular}[c]{@{}c@{}}GitHub\\ (XGB)\end{tabular}                                  & 0.72        & 0.58        & 0.64       & 0.66        & 0.82       & 0.73       & 0.63        & 0.54        & 0.58       & 0.67        & 0.65        & 0.65        & 0.67                                \\
                                                                                         & \begin{tabular}[c]{@{}c@{}}Jira\\ (LightGBM)\end{tabular}                               & 0.72        & 0.83        & 0.77       & 0.88        & 0.86       & 0.87       & 0.69        & 0.59        & 0.64       & 0.76        & 0.76        & 0.6         & 0.82                                \\ \cline{2-15} 
\multirow{2}{*}{\begin{tabular}[c]{@{}c@{}}IBM Watson \\ AutoAI\end{tabular}}            & \begin{tabular}[c]{@{}c@{}}GitHub\\ (XGB)\end{tabular}                                  & 0.39        & 0.62        & 0.48       & 0.80        & 0.60       & 0.69       & 0.46        & 0.53        & 0.49       & 0.55        & 0.58        & 0.55        & 0.58                                \\
                                                                                         & \begin{tabular}[c]{@{}c@{}}Jira\\ (XGB)\end{tabular}                                    & 0.55        & 0.73        & 0.63       & 0.93        & 0.77       & 0.84       & 0.24        & 0.69        & 0.35       & 0.57        & 0.73        & 0.61        & 0.76                                \\ \cline{2-15} 
\multirow{2}{*}{\begin{tabular}[c]{@{}c@{}}Ludwig \\ AutoML\end{tabular}}                & \begin{tabular}[c]{@{}c@{}}GitHub\\ (Nueral Net.)\end{tabular}                          & 0.90        & 0.89        & 0.90       & 0.85        & 0.89       & 0.87       & 0.86        & 0.81        & 0.83       & 0.87        & 0.86        & 0.87        & 0.87                                \\
                                                                                         & \begin{tabular}[c]{@{}c@{}}Jira\\ (Nueral Net.)\end{tabular}                            & 0.71        & 0.79        & 0.75       & 0.87        & 0.88       & 0.88       & 0.88        & 0.64        & 0.74       & 0.82        & 0.77        & 0.79        & 0.83                                \\ \cline{2-15} 
\multirow{2}{*}{\begin{tabular}[c]{@{}c@{}}MS Azure \\ AutoML\end{tabular}}              & \begin{tabular}[c]{@{}c@{}}GitHub\\ (XGB)\end{tabular}                                  & 0.88        & 0.87        & 0.88       & 0.,87       & 0.88       & 0.88       & 0.84        & 0.84        & 0.84       & 0.87        & 0.86        & 0.86        & 0.87                                \\
                                                                                         & \begin{tabular}[c]{@{}c@{}}Jira\\ (LightGBM)\end{tabular}                               & 0.83        & 0.75        & 0.78       & 0.88        & 0.92       & 0.90       & 0.77        & 0.71        & 0.74       & 0.83        & 0.80        & 0.81        & 0.86                                \\ \hline
\multirow{2}{*}{\fullcite{Novielli2020}}                                                 & \begin{tabular}[c]{@{}c@{}}GitHub\\ (Senti4SD)\end{tabular}                                                                                   & 0.95        & 0.91        & 0.93       & 0.90        & 0.93       & 0.92       & 0.92        & 0.90        & 0.91       & \textbf{0.92}        & 0.91        & \textbf{0.92}        & \textbf{0.92}                                \\
                                                                                         & \begin{tabular}[c]{@{}c@{}}Jira\\ (SentiCR)\end{tabular}                                                                                     & 0.79        & 0.88        & 0.83       & 0.81        & 0.91       & 0.63       & 0.80        & 0.89        & 0.72       & 0.83        & 0.78        & 0.80        & 0.86                                \\ \cline{2-15} 
\fullcite{Zhang2020}                                                                     & \begin{tabular}[c]{@{}c@{}}GitHub\\ (RoBERTa)\end{tabular}                                                                                   & 0.93        & 0.96        & 0.94       & 0.91        & 0.92       & 0.92       & 0.93        & 0.89        & 0.91       & \textbf{0.92}        & \textbf{0.92}        & \textbf{0.92}        & \textbf{0.92}                                \\ \hline
\end{tabular}
\vspace{-2mm}
\end{table*}

\section{Results}\label{sec:results}

\subsection{Quantitative Analysis}\label{sec:res-quantitative}

In this section, we report the performance of the twelve \aml\ tools on the two sentiment analysis datasets described earlier. 
Results are reported in Table~\ref{tab:results}.
For each dataset, we highlight in bold the best performance in terms of the two main metrics (i.e., macro- and micro-averaged F1-scores). 

\textbf{GitHub dataset}. The GitHub dataset is the more balanced of the two analyzed in this benchmark.
Previous work by \fullcite{Novielli2020} and \fullcite{Zhang2020} report performance in terms of micro- and macro-average F1-score between 0.91 and 0.92, which we use as a reference for state-of-the-art performance on the dataset.
In the case of \fullcite{Zhang2020}, the best performance reported was obtained using the pre-trained transformer model RoBERTa; in the case of \fullcite{Novielli2020}, the best results reported refer to the Senti4SD tool.
DataRobot AI Cloud obtained the  best results among the \aml\ solutions. It performed as good as the pre-trained transformers assessed in~\fullcite{Zhang2020} and obtained a slightly smaller macro-averaged Recall as compared to the Senti4SD tool benchmarked in~\fullcite{Novielli2020} (0.91 vs. 0.92).
The second best-performing tool is AutoKeras, which obtained a score of 0.91 on all the metrics.
The worst results were obtained by IBM Watson AutoAI, which obtained 0.58 micro-average and macro-averaged scores ranging between 0.55 and 0.58, and H20 Driverless AI, which obtained 0.67 micro-average and macro-averaged scores ranging between 0.65 and 0.67.
Regarding the remaining tools, we observe a subset of solutions whose micro-average F1-scores range between 0.70 and 0.75 (i.e., AutoGluon, Amazon SageMaker Autopilot, and BigML) and another subset consisting of tools with a performance between 0.81 and 0.89 (i.e., AutoGoal, Ludwig AutoML, MS Azure AutoML, and Google Vertex AI).

\textbf{Jira dataset}. The Jira dataset is less balanced than the GitHub one, which might explain the worse performance in general.
\fullcite{Novielli2020} report a micro-average score of .86 and macro-averaged scores between 0.78 and 0.83.
AutoKeras, BigML, and Google Vertex AI outperform by a small margin the SE-specific tools as they reach a micro-average score of 0.87, followed by MS Azure AutoML and Amazon SageMaker Autopilot (0.86). 
The performances of the \aml\ tools in terms of macro-averaged Precision, Recall, and F1, although varied, are equal to or better than (values between 0.80 and 0.85) the best performance reported in~\fullcite{Novielli2020} (between 0.80 and 0.83).
For this dataset, the worst performances are those achieved by IBM Watson AutoAI (macro-averaged scores between 0.57 and 0.73, micro-average 0.76) and AutoGoal (\tildex{0.70} macro-average scores, 0.75 micro-average). 

Overall, we found that AutoKeras (open source) and DataRobot AI Cloud (commercial) perform best on both datasets; both \aml\ tools obtained the best scores with neural network models, reaching a performance that is on par with SE-specific research tools for the GitHub dataset and slightly better for the Jira dataset.
Other solutions such as Google Vertex AI and MS Azure AutoML (commercial) and Ludwig AutoML (open source) perform well on average on both datasets; these solutions built either a neural network or a light/extreme gradient boosting machine model.
IBM Watson AutoAI is the \aml\ tool that performed consistently worse on both datasets.


\subsection{Qualitative Analysis}\label{sec:res-qualitative}

\subsubsection{Survey}
We received 51 answers. 
However, 6 came from respondents who reported having heard of \aml\ but never used it; hence, we excluded them and the analysis reported next focuses on the remaining $n=45$ subjects.

\textbf{Participants' background}.
Here we check whether sampled participants are representative of the target population (i.e., software engineers working on AI/ML projects and having experience with \aml).
The subjects identify themselves mostly as male (39, 87\%) and have ages in the range of 26-35 (25, 56\%) and 36-45 (14, 31\%).
They come from different countries, in particular Italy (9, 20\%), Brazil (8, 18\%), the United States (7, 16\%), and Canada (6, 13\%). 

The respondents reported having mostly a background in CS, namely a PhD (16, 36\%), MSc (10, 22\%), or BSc (6, 13\%); the remaining respondents have an MSc in AI/Computational Intelligence (7, 16\%), BSc in Statistics (4, 9\%), and PhD in Physics (2, 4\%).
About half of them have an overall working experience between 5-10 years (17, 26\%) or more than 10 years (14, 22\%).
On a five-point Likert scale (1=\textit{Not experienced at all}, 5=\textit{Very experienced}) in SE practices (e.g., designing, implementing, testing, and maintaining complex software applications), most of them reported a high (18, 40\%) or very high (15, 33\%) level of expertise.
Regarding AI/ML specifically, the respondents mostly reported having some experience (i.e., 3-4 years, 20, 45\%) or considerable experience (i.e., 5-10 years, 15, 33\%), with the remaining ones (10, 22\%) having limited experience (1-2 years).
Most of the respondents identify their work activities (answers not mutually exclusive) as pertaining to the positions of ML engineer (34), followed by MLOps (15), Researcher (7), and Data Scientist (7) in  either a tech company (67\%) or non-commercial/academic research lab (24\%).

The sizes of the companies for which respondents work are varied; about half of the respondents are working for either very large companies (10,000+ employees, 27\%) or very small ones (10-50 employees, 25\%).
Most of the subjects (17, 38\%) work in teams of 6 to 9 members; very small (2-3 members) and very large (16+) teams are also common (27\% and 22\% of the responses, respectively).
This shows that AutoML is a technology appealing to all types of companies, regardless of their size.
Half of the respondents report that \tildex{35-60}\% of their team members (median=3) have a strong SE background.
Most of the teams they work in are either well-balanced, interdisciplinary teams (23, 51\%) or teams made of mostly AI/ML people (17, 38\%); less common are the cases where most team members are software engineers (5, 11\%).

Respondents mostly work on projects in the domain of healthcare (19), banking \& insurance (16), and telecommunication (15),  where AI/ML models are either part of a product (10, 22\%) or integrated into a larger, production-ready software system (25, 56\%).
The primary use cases for AI/ML projects are classification (15), prediction (13), forecasting (12), risk evaluation (11), and anomaly detection (10).  
The typical lengths of projects are 4-6 months (14, 31\%) and 7-12 months (13, 29\%); less common are projects longer than one year (11, 24\%) or taking only 2-3 months (7, 16\%).
With respect to the AI/ML reference workflow in Fig.~\ref{fig:ml-workflow}, respondents report participating mostly in the analysis stage, i.e., feature engineering (39), model evaluation (34), and training (30).
Dissemination activities are also common, i.e., model deployment (22) and monitoring (13), whereas the involvement in preparation activities, i.e., data cleaning (6), labeling (7), and collection (8), is far less common.
Only one respondent reported being involved in the whole end-to-end process.

\textbf{Usage and features of \aml}.
Overall, the survey reveals that among software engineers there is considerable interest in \aml\ solutions, whether applied on a regular basis (15, 33\%) or on select projects (23, 51\%), with the remaining respondents (7, 16\%) reporting having used it at least  in internal projects.
They report using or having experimented with both proprietary cloud solutions (in order of popularity MS Azure \aml, Amazon SageMaker AutoPilot, and IBM Watson AutoAI) and open-source tools (AutoKeras, TPOT, auto-sklearn, nni, and Alteryx); the cited reasons are different: for proprietary tools, the reasons are  vendor lock-in and clients' trust in renowned companies; for open-source solutions, it is the flexibility in adapting to custom workflows.

\begin{figure}
    \centering
    \vspace{-6mm}
    \includegraphics[scale=0.35]{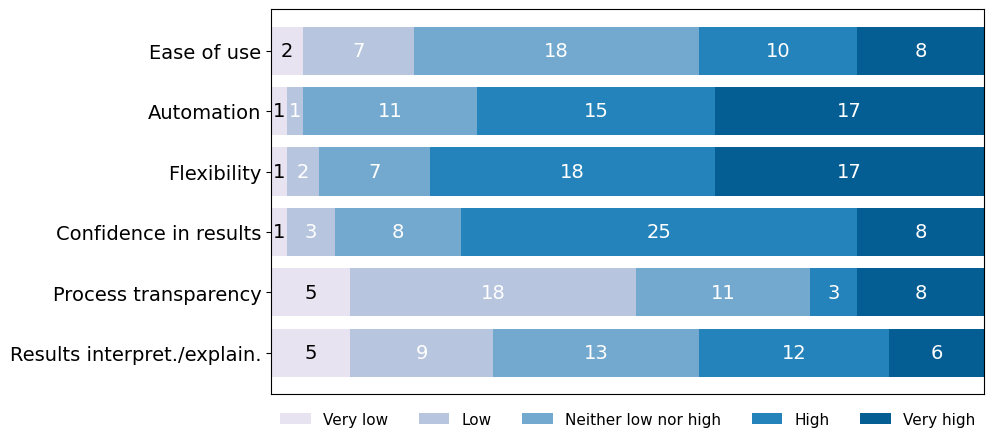}
    \vspace{-7mm}
    \caption{Features more relevant for choosing an \aml\ solution.}
    \vspace{-6mm}
    \label{fig:likert}
\end{figure}

Table~\ref{tab:tools} shows that the features in the  analysis stage (feature engineering plus model training and evaluation) are the most widely supported among the reviewed tools.
Therefore, it is not surprising that the questionnaire respondents report mostly using \aml\ for model training (45, 100\%), model evaluation (33, 73\%), and feature engineering (20, 44\%).
Notably, the use of \aml\ for model deployment and monitoring (17, 20\%) is considerably more common than data preparation activities (5, 10\%).
None of the participants reported using \aml\ in \textit{full automation} mode, with humans out of the loop.
Most respondents (22, 49\%) use \aml\ in \textit{human-directed automation} mode, whereby they can select methods and parameters before execution; 
the remaining ones use it in \textit{system-suggested automation} mode (12, 27\%, i.e., the system decides on parameters/methods, and human approval is requested before execution) and \textit{system-directed automation} (9, 20\%, i.e., the system executes automatically, and humans are in the loop and can intervene anytime). 
However, when choosing an \aml\ solution (Fig.~\ref{fig:likert}), respondents value the flexibility to customize and adapt the workflow to their needs (78\% either \textit{very high} or \textit{high}) and automation capabilities (71\% either \textit{very high} or \textit{high}) more than ease of use (40\%).
They report having a substantially high level of confidence in the results obtained using \aml\ tools (73\% either \textit{very high} or \textit{high}).
In addition, few respondents find the process to be transparent (51\% either \textit{very low} or \textit{low}) and they are also somewhat conflicted when it comes to the interpretability/explainability of results obtained via \aml.

\subsubsection{Emerging Themes}
Five of the eight respondents who had given their consent agreed to a follow-up interview. 
All of them opted for conducting it over email. 
As such, we defined and emailed four main questions; then, they answered by email and we replied back to request further clarification when needed.
Here we report the analysis of the answers to both the open-ended questions from the survey and the follow-up interviews (see the supplemental material).
We followed a sequential explanatory strategy whereby we first performed an open coding of the responses to the open-ended survey questions to identify common themes; then, we formulated  \textit{ad hoc} questions for the interviewees to gather further insights.
We were able to identify three emerging themes, which we illustrate next by providing relevant excerpts.
We use the code $P_x$ to report quotes from survey participants and $PI_y$ to distinguish those who also agreed to the follow-up interview (Table \ref{tab:interviewees} provides an overview).

\textbf{AI/ML team members wear different hats}. 
Working on AI/ML projects is a team sport involving different activities and skills. 
Still, we found that in these teams, member roles tend to be  blurred; 
especially in smaller teams, it is not unusual for surveyed participants to perform activities belonging to different stages of the AI/ML workflow, in particular, analysis and dissemination; (only a few reported participating in data preparation activities).
This has a considerable impact on expectations and perceived usefulness of \aml.

Albeit rarely involved in it, participants are skeptical that \aml\ might ever automate data preparation activities ($P_2$, $P_{11}$, $P_{21}$, $P_{33}$):

\begin{displayquote}
``\textit{I'm not involved, not often at least, but data people in my teams tell me that [\aml] support is garbage. I don't think it's even possible to automate it. I wish, but it's too domain dependent}...'' ($PI_{12}$)
\end{displayquote}

The analysis stage is the one that \aml\ appears to support better and with which most of the respondents reported to be satisfied, although some ($P_{14}$, $P_{31}$) feel that ``\textit{feature engineering might be improved''} ($P_{8}$).
The dissemination stage is the one that has more room for improvement. 
Our analysis suggests that this deficiency might be because ML models do not always end up being deployed into production systems (``\textit{sometimes our customers just want us to uncover insights about features and relationships between them},'' $P_{26}$).
Still, several respondents ($P_{4}$, $PI_{12}$, $P_{16}$, $PI_{23}$) were vocal about wishing to see better automation support for model deployment and monitoring:
\begin{displayquote}
``\textit{Cloud solutions force you to use their infrastructure. [They] don't help us deploy where we want. They are not flexible. As an ML engineer, I want to choose where to deploy and how to serve models. We have our own infrastructure and monitoring dashboard that we'd like to keep using}.'' ($PI_{23}$)
\end{displayquote}
\begin{displayquote}
``\textit{[\aml] is just a commercial name for a glorified version of HPO and CASH tools. Automation right now happens only at the analysis stage [...] The other activities [are] not first-class citizens}.'' ($P_{16}$)
\end{displayquote}

\begin{table}[t]
\centering
\vspace{-6mm}
\caption{Overview of interviewees.}
\vspace{-2mm}
\label{tab:interviewees}
\begin{tabular}{ccllc}
\hline
\textbf{ID} & \textbf{Gender} & \textbf{Country} & \textbf{Domain}                                              & \textbf{Job description}                                                  \\ \hline
$PI_{12}$     & Male            & USA              & Healthcare                                                   & ML engineer                                                               \\
$PI_{23}$     & Male            & Germany          & Telecom                                                     & ML engineer    \\
$PI_{29}$     & Male            & USA              & Banking                                                      & ML engineer                                                               \\
$PI_{40}$     & Male            & Canada           & Healthcare                                                   & ML engineer \\
$PI_{44}$     & Male            & Sweden           & Banking, insurance & MLOps     \\ \hline
\end{tabular}
\vspace{-5mm}
\end{table}

\textbf{ML-as-a-Service vs. in-house solutions}. Another theme that emerged is that cloud solutions that offer ML as a service might be more appealing to smaller teams, where roles tend to be blurred:
\begin{displayquote}
``\textit{These MLaaS commercial tools force you to embrace their solution [end-to-end]. Management and customers trust them because they are sold by well-known vendors. But having worked in teams of different sizes on different projects, I feel that those are more useful in scenarios where they ask you to build a fast PoC [...] to show the customer how good the model can be and then give the data scientists a baseline to work on if the project is greenlit}.'' ($PI_{40}$)
\end{displayquote}
This lack of flexibility is not appreciated by respondents working in larger teams with more clearly defined roles. They would want to see \aml\ not as one tool, but either as  a collection of (``\textit{smaller scope tools, something to plug into your own workflow}'', $P_{17}$) or as a `generator' of custom, project-specific workflows rather than an `automator':
\begin{displayquote}
``\textit{We tinkered with many \aml\ alternatives before deciding to build one around OSS libraries and tools we already used and liked. It's actually more of an \aml\ workflow generator}.'' 
($PI_{23}$)
\end{displayquote}

\textbf{Tension between automation desire and user agency}. 
The analysis revealed the existence of a contrast among respondents between the need for human control and a higher desire for automation.
In particular, our findings suggest that the desire for higher automation varies with the stages of the ML workflow in which participants are involved and also clashes with the difficulty of actually achieving it.
As reported earlier, data preparation is especially domain-dependent, and albeit those involved in data preparation do wish for more automation, automating domain knowledge elicitation remains a prohibitive task.
The analysis stage is the one where the highest level of automation is achieved ($P_{9}$, $P_{16}$, $P_{25}$, $P_{28}$); yet, respondents are also aware of the need for rather using \aml\ in ``\textit{cruise control mode}'' ($P_{25}$), where humans ``\textit{stay in the loop to oversee and steer the system, and possibly get explanations if need be}'' ($PI_{29}$), due to regulatory and organizational requirements.
Given the large presence of ML engineers and MLOps among respondents, it is not surprising that we found participants to be eager to achieve better support for automating the dissemination stage ($PI_{12}$, $P_{16}$, $PI_{23}$, $PI_{44}$), where human oversight is perceived both less problematic and intrinsic to monitoring:

\begin{displayquote}
``\textit{Full automation is problematic, with ethical concerns and regulations [...]. So you want control when you need to understand data, understand models. But there are no such issues with what you call dissemination. Better (if not fully) automated deployment would be amazing... there user control comes just with monitoring}.'' ($PI_{44}$)
\end{displayquote}

\begin{displayquote}
``\textit{Dissemination [is poorly supported] because most of the models rarely get into production. However, for us ML engineers and MLOps there is a lot of code generation involved to integrate them into the final system and deploy the whole thing. A lot of repetition is there that could be  automated with \aml}.'' ($PI_{12}$)
\end{displayquote}

\section{Discussion}\label{sec:discussion}
In this study, we investigated two research questions, namely RQ1 (\aml\ tools benchmark) and RQ2 (software engineers' perspective on using \aml).
RQ1 is answered in Section~\ref{sec:discrq1}, whereas the findings related to RQ2 are discussed in Sections~\ref{sec:discrq2a} and \ref{sec:discrq2b}.

\vspace{-1mm}
\subsection{\aml\ as a Solution to Augment SE Research}\label{sec:discrq1}
Our benchmark revealed that, on average, AutoKeras (open source) and DataRobot AI Cloud (commercial) are the best performing \aml\ solutions for sentiment analysis on the two datasets analyzed.
These tools perform on par with both the pre-trained transformer-based models and SE-specific research tools in the case of the GitHub dataset, and slightly better in the case of the Jira dataset.
    
Although there is no gold standard or concrete threshold of various evaluation metrics to decide whether a SE-specific sentiment analysis tool can be put into real use, our experiment results show that the use of \aml\ is practical for sentiment analysis tasks in SE.
It is worth noticing that the performance of the best \aml\ tool is on par with state-of-the-art results and SE-specific research tools built by human experts, who oversaw all the steps within the development pipeline.
On the contrary, \fullcite{Zoller2021} compared the performance of a set of five \aml\ tools (different than those benchmarked in this study) against a couple of Kaggle notebooks containing the best-performing pipelines developed by human experts on a couple of public datasets;
they found the models built by human experts to outperform the \aml\ tools, which in general achieved mediocre performance in comparison. 
As such, although the mileage may vary with the specific tool and tasks other than sentiment analysis, we suggest  researchers consider the use of \aml\ solutions to train initial models as baselines or starting points for future work. 
Finally, we note that our assessment here pertains only to the analysis stage. 
We ignored the preparation and dissemination stages, which we discuss next.

\textbf{Implications}. 
Echoing the words of \fullcite{Wang2019} and \fullcite{Crisan2021}, our findings encourage SE researchers to use \aml\ for \textit{augmenting} their research activities; in particular, by automating model building and evaluation, they can spend their (limited) resources on time-consuming activities such data collection and cleaning, which still heavily depend on the elicitation of domain knowledge and human expertise.


\vspace{-0.5mm}
\subsection{Uneven Automation Through the AI/ML Workflow}\label{sec:discrq2a}

The progress in AI/ML research in the last few years has been enormous and we now are witnessing a fundamental shift in software engineering where machine learning becomes the new software, powered by big data and computing infrastructure. 
Much like cloud resources have made it easy to access high-performing hardware, resources like the Hugging Face model registry, 
Large Language Models (LLMs), and \aml\ are commoditizing ML models, which become either readily available to developers or extraordinarily easier to build and fine-tune. 
As a result, a project's success now depends on the (quality of) data rather than models.

Despite AI and ML being more and more data-centric, we found out that \aml\ solutions currently provide little or no support to the data preparation stage.
As $P_2$ put it succinctly, ``\textit{we just don't use \aml\ for data preparation}.'' 
While participants remain generally skeptical that data preparation can be fully automated due to its heavy domain dependence and despite not usually being involved in such activities, they were also very vocal about the benefits that better tool support would bring to the AI/ML teams.
Similarly, participants reported among their desiderata better support for dissemination -- deployment in particular, since monitoring is seen as the means to enforce user oversight. 
Even if not completely automated, our respondents wish for an assisted and flexible solution that saves them from the tedium of writing boilerplate integration code and allows them to customize the deployment destination.

Our finding of uneven automation support of workflow stages adds a potential explanation for the low adoption of \aml\ practices reported in previous surveys also conducted in the context of software engineering for ML~\cite{Serban2020,vanderBlom2021}.

\textbf{Implications.} The automation of the analysis stage appears to have been tackled albeit still needing some improvements (e.g., explainability, feature engineering).
To distinguish themselves from competitors, \aml\ tool designers must focus on providing better assistance to data preparation and dissemination stages.

\vspace{-1.0mm}
\subsection{\aml\ is a Misnomer (for non-Data Scientists)}\label{sec:discrq2b}
In their studies on the use of \aml\ among data scientists, \fullcite{Wang2019} and \fullcite{Crisan2021} reported two common usage scenarios: (i) rapid model prototyping and (ii) data science democratization.
Our qualitative analysis revealed that while the first  scenario is also common among our participants, the second one is not, arguably because of the different target population: being versed in ML/software engineering practices and with coding abilities, our study participants do not see value in using \aml\ as a means to lower the entry barriers for data science. 

Consistently with previous research that already reported on the lack of customizability in existing \aml\ solutions~\cite{Xin2021, Sun2023}, our findings add further evidence that experienced ML engineers/MLOps are more interested in flexibility and control than ease of use.
As such, the desire for full automation appears to be inversely proportional to AI/ML expertise, and, given that ML engineers/MLOps rather look at these solutions as generators of custom AI/ML workflows, we argue that \aml\ is currently a misnomer since the promise of automation is only fulfilled for data scientists at the model analysis stage.

\textbf{Implications}.
Echoing the words of~\fullcite{Xin2021}, to appeal to a broader audience, \aml\ tool builders should refrain from using commercial names suggesting human-out-of-the-loop, full-automation capabilities (e.g., autopilot, driverless) given that full-automation has proven infeasible so far and only appealing to non-experts who want a one-click model building solution, while also being in contrast with the recent EU ethics regulations, 
which require AI/ML systems to empower humans  
through human-in-the-loop approaches.

Because AI/ML workflow activities are many and complex, \aml\ users are also diverse, with different backgrounds and contrasting desires for control and automation.
We argue that \aml\ product designers should consider using personas~\cite{Pruitt2003personas}, a well-known product design technique that allows to model archetypal users and better capture their goals, needs, and frustrations.
In the case of \aml, it is easy to envision at least three types of users, one for each of the three stages in the AI/ML workflow because, even though the roles may be blurred in teams, especially small, the user needs and focus at each stage are clearly defined and different.

\vspace{-0.5mm}
\subsection{Limitations}
This study has a few limitations that can inform future research.
Regarding the benchmark, some of the existing OSS \aml\ solutions were excluded because, albeit supporting text features, their documentation was not providing clear examples similar to the chosen experimental task.
While we acknowledge this as a potential limitation of the tool sampling process, we also point out that the focus here was not the completeness of the benchmark. 
In fact, our results already show that on-premise OSS solutions like AutoKeras can keep up with commercial cloud solutions and automatically generate models with similar if not better performance.
Regarding the choice of sentiment analysis as the experimental task, we acknowledge a partial answer to RQ1.
There is indeed a variety of other SE-specific classification tasks (e.g., defect prediction, bug classification). 
However, 
we initially opted for a task whose execution would require expertise that is typically outside of the knowledge domain of software engineers (i.e., NLP), thus showing how \aml\ can be a Swiss knife for building  ML models even on limited skills.
Regarding the survey, albeit we found that participants come from a variety of backgrounds and use \aml\ to different extents, the voluntary participation may have resulted in self-selection bias, with those responding being more active with or positive towards  \aml\ than those who did not.

\section{Conclusions}\label{sec:conclusions}

In this study, we investigated the use of \aml\ for data-driven software engineering. 
Our findings suggest that \aml\ promises to generate (classification) models that outperform those trained and optimized by researchers in the SE domain, making it a valuable aid for automating the heavy lifting in model analysis. 
However, we also found that the currently available \aml\ solutions do not equally support automation across all stages of the ML development workflow, with data/ML engineers and MLOps currently being second-class citizens. 
Overall, our study contributed insights for the SE research community and tool builders on how to use \aml\ technologies and improve the design of its next generation.

\vspace{-1mm}
\section*{Acknowledgment}

The authors would like to thank the survey participants. This research was co-funded by projects DARE (PNC0000002, CUP: B53C22006420001), FAIR (PE00000013, CUP: H97G22000210007), and SERICS (PE0000014, CUP: H93C22000620001).

\bibliographystyle{IEEEtranN}
\bibliography{IEEEabrv, references}

\begin{thebibliography}{37}
\providecommand{\natexlab}[1]{#1}
\providecommand{\url}[1]{#1}
\csname url@samestyle\endcsname
\providecommand{\newblock}{\relax}
\providecommand{\bibinfo}[2]{#2}
\providecommand{\BIBentrySTDinterwordspacing}{\spaceskip=0pt\relax}
\providecommand{\BIBentryALTinterwordstretchfactor}{4}
\providecommand{\BIBentryALTinterwordspacing}{\spaceskip=\fontdimen2\font plus
\BIBentryALTinterwordstretchfactor\fontdimen3\font minus
  \fontdimen4\font\relax}
\providecommand{\BIBforeignlanguage}[2]{{%
\expandafter\ifx\csname l@#1\endcsname\relax
\typeout{** WARNING: IEEEtranN.bst: No hyphenation pattern has been}%
\typeout{** loaded for the language `#1'. Using the pattern for}%
\typeout{** the default language instead.}%
\else
\language=\csname l@#1\endcsname
\fi
#2}}
\providecommand{\BIBdecl}{\relax}
\BIBdecl

\bibitem[Jordan and Mitchell(2015)]{Jordan2015}
M.~I. Jordan and T.~M. Mitchell, ``Machine learning: Trends, perspectives, and
  prospects,'' \emph{Science}, vol. 349, no. 6245, pp. 255--260, 2015.

\bibitem[Luckow et~al.(2018)Luckow, Kennedy, Ziolkowski, Djerekarov, Cook,
  Duffy, Schleiss, Vorster, Weill, Kulshrestha, et~al.]{Luckow2018}
A.~Luckow, K.~Kennedy, M.~Ziolkowski, E.~Djerekarov, M.~Cook, E.~Duffy,
  M.~Schleiss, B.~Vorster, E.~Weill, A.~Kulshrestha \emph{et~al.}, ``Artificial
  intelligence and deep learning applications for automotive manufacturing,''
  in \emph{2018 IEEE Int'l Conf. on Big Data (Big Data)}.\hskip 1em plus 0.5em
  minus 0.4em\relax IEEE, 2018, pp. 3144--3152.

\bibitem[Markow et~al.()Markow, Braganza, Taska, Miller, and
  Hughes]{Markow2017}
\BIBentryALTinterwordspacing
W.~Markow, S.~Braganza, B.~Taska, S.~M. Miller, and D.~Hughes, ``The quant
  crunch: How the demand for data science skills is disrupting the job
  market.'' [Online]. Available:
  \url{https://www.ibm.com/downloads/cas/3RL3VXGA}
\BIBentrySTDinterwordspacing

\bibitem[Chakravorti et~al.()Chakravorti, Bhalla, Shankar~Chaturvedi, and
  Filipovic]{Bhaskar2021}
\BIBentryALTinterwordspacing
B.~Chakravorti, A.~Bhalla, R.~Shankar~Chaturvedi, and C.~Filipovic, ``50 global
  hubs for top ai talent.'' [Online]. Available:
  \url{https://hbr.org/2021/12/50-global-hubs-for-top-ai-talent}
\BIBentrySTDinterwordspacing

\bibitem[Zhang et~al.(2021)Zhang, Mishra, Brynjolfsson, Etchemendy, Ganguli,
  Grosz, Lyons, Manyika, Niebles, Sellitto, et~al.]{Zhang2021}
D.~Zhang, S.~Mishra, E.~Brynjolfsson, J.~Etchemendy, D.~Ganguli, B.~Grosz,
  T.~Lyons, J.~Manyika, J.~C. Niebles, M.~Sellitto \emph{et~al.}, ``The ai
  index 2021 annual report,'' \emph{arXiv preprint arXiv:2103.06312}, 2021.

\bibitem[Hutter et~al.(2019)Hutter, Kotthoff, and Vanschoren]{Hutter2019}
F.~Hutter, L.~Kotthoff, and J.~Vanschoren, \emph{Automated machine learning:
  methods, systems, challenges}.\hskip 1em plus 0.5em minus 0.4em\relax
  Springer Nature, 2019.

\bibitem[Tanaka et~al.(2019)Tanaka, Monden, and Yücel]{Tanaka2019}
K.~Tanaka, A.~Monden, and Z.~Yücel, ``Prediction of software defects using
  automated machine learning,'' in \emph{2019 20th IEEE/ACIS International
  Conference on Software Engineering, Artificial Intelligence, Networking and
  Parallel/Distributed Computing (SNPD)}, 2019, pp. 490--494.

\bibitem[Vanschoren et~al.(2014)Vanschoren, Van~Rijn, Bischl, and
  Torgo]{Vanschoren2014}
J.~Vanschoren, J.~N. Van~Rijn, B.~Bischl, and L.~Torgo, ``Openml: networked
  science in machine learning,'' \emph{ACM SIGKDD Explorations Newsletter},
  vol.~15, no.~2, pp. 49--60, 2014.

\bibitem[Van~der Blom et~al.(2021)Van~der Blom, Serban, Hoos, and
  Visser]{vanderBlom2021}
K.~Van~der Blom, A.~Serban, H.~Hoos, and J.~Visser, ``{AutoML Adoption in ML
  Software},'' \emph{8th ICML Workshop on Automated Machine Learning}, 2021.

\bibitem[Wang et~al.(2019)Wang, Weisz, Muller, Ram, Geyer, Dugan, Tausczik,
  Samulowitz, and Gray]{Wang2019}
D.~Wang, J.~D. Weisz, M.~Muller, P.~Ram, W.~Geyer, C.~Dugan, Y.~Tausczik,
  H.~Samulowitz, and A.~Gray, ``{Human-AI Collaboration in Data Science:
  Exploring Data Scientists’ Perceptions of Automated AI},'' \emph{Proc. of
  ACM on Human-Computer Interaction}, vol.~3, no. CSCW, pp. 1--24, nov 2019.

\bibitem[Sato et~al.(2019)Sato, Wider, and Windheuser]{Sato2019}
\BIBentryALTinterwordspacing
D.~Sato, A.~Wider, and C.~Windheuser, ``Continuous delivery for machine
  learning,'' Sep 2019. [Online]. Available:
  \url{https://martinfowler.com/articles/cd4ml.html}
\BIBentrySTDinterwordspacing

\bibitem[Menzies(2019)]{Menzies2019}
T.~Menzies, ``The five laws of se for ai,'' \emph{IEEE Software}, vol.~37,
  no.~1, pp. 81--85, 2019.

\bibitem[Wang et~al.(2023)Wang, Chen, and Zhou]{Wang2023}
C.~Wang, Z.~Chen, and M.~Zhou, ``{AutoML from Software Engineering Perspective:
  Landscapes and Challenges},'' in \emph{2023 Proc. of 20th Int.l Conf. on
  Mining Software Repositories (MSR)}, may 2023.

\bibitem[Majidi et~al.(2022)Majidi, Openja, Khomh, and Li]{Majidi2022}
\BIBentryALTinterwordspacing
F.~Majidi, M.~Openja, F.~Khomh, and H.~Li, ``An empirical study on the usage of
  automated machine learning tools,'' in \emph{{IEEE} International Conference
  on Software Maintenance and Evolution, {ICSME} 2022, Limassol, Cyprus,
  October 3-7, 2022}.\hskip 1em plus 0.5em minus 0.4em\relax {IEEE}, 2022, pp.
  59--70. [Online]. Available:
  \url{https://doi.org/10.1109/ICSME55016.2022.00014}
\BIBentrySTDinterwordspacing

\bibitem[Gijsbers et~al.(2019)Gijsbers, LeDell, Thomas, Poirier, Bischl, and
  Vanschoren]{Gijsbers2019}
P.~Gijsbers, E.~LeDell, J.~Thomas, S.~Poirier, B.~Bischl, and J.~Vanschoren,
  ``{An Open Source AutoML Benchmark},'' in \emph{6th ICML Workshop on
  Automated Machine Learning}, jul 2019, pp. 1--8.

\bibitem[Guyon et~al.(2019)Guyon, Sun-Hosoya, Boull{\'{e}}, Escalante,
  Escalera, Liu, Jajetic, Ray, Saeed, Sebag, Statnikov, Tu, and
  Viegas]{Guyon2019}
I.~Guyon, L.~Sun-Hosoya, M.~Boull{\'{e}}, H.~J. Escalante, S.~Escalera, Z.~Liu,
  D.~Jajetic, B.~Ray, M.~Saeed, M.~Sebag, A.~Statnikov, W.-W. Tu, and
  E.~Viegas, ``{Analysis of the AutoML Challenge Series 2015–2018},'' in
  \emph{Automated Machine Learning}, V.~J. {Hutter F., Kotthoff L.}, Ed.\hskip
  1em plus 0.5em minus 0.4em\relax Springer, 2019, ch.~10, pp. 177--219.

\bibitem[Z{\"{o}}ller and Huber(2021)]{Zoller2021}
M.-A. Z{\"{o}}ller and M.~F. Huber, ``{Benchmark and Survey of Automated
  Machine Learning Frameworks},'' \emph{Journal of Artificial Intelligence
  Research}, vol.~70, pp. 409--472, jan 2021.

\bibitem[Paldino et~al.(2021)Paldino, {De Stefani}, {De Caro}, and
  Bontempi]{Paldino2021}
G.~M. Paldino, J.~{De Stefani}, F.~{De Caro}, and G.~Bontempi, ``{Does AutoML
  Outperform Naive Forecasting?}'' \emph{Engineering Proceedings}, vol.~5,
  no.~1, p.~36, jul 2021.

\bibitem[Devlin et~al.(2018)Devlin, Chang, Lee, and Toutanova]{Devlin2018}
J.~Devlin, M.-W. Chang, K.~Lee, and K.~Toutanova, ``Bert: Pre-training of deep
  bidirectional transformers for language understanding,'' \emph{arXiv preprint
  arXiv:1810.04805}, 2018.

\bibitem[Epperson et~al.(2022)Epperson, Wang, DeLine, and
  Drucker]{Epperson2022}
W.~Epperson, A.~Y. Wang, R.~DeLine, and S.~M. Drucker, ``Strategies for reuse
  and sharing among data scientists in software teams,'' 2022.

\bibitem[Amershi et~al.(2019)Amershi, Begel, Bird, DeLine, Gall, Kamar,
  Nagappan, Nushi, and Zimmermann]{Amershi2019}
S.~Amershi, A.~Begel, C.~Bird, R.~DeLine, H.~Gall, E.~Kamar, N.~Nagappan,
  B.~Nushi, and T.~Zimmermann, ``Software engineering for machine learning: A
  case study,'' in \emph{2019 IEEE/ACM 41st Int'l Conf. on Software
  Engineering: Software Engineering in Practice (ICSE-SEIP)}.\hskip 1em plus
  0.5em minus 0.4em\relax IEEE, 2019, pp. 291--300.

\bibitem[Hanussek et~al.(2020)Hanussek, Blohm, and Kintz]{Hanussek2020}
M.~Hanussek, M.~Blohm, and M.~Kintz, ``{Can AutoML outperform humans? An
  evaluation on popular OpenML datasets using AutoML Benchmark},'' in
  \emph{2020 2nd Int'l Conf. on Artificial Intelligence, Robotics and
  Control}.\hskip 1em plus 0.5em minus 0.4em\relax New York, NY, USA: ACM, dec
  2020, pp. 29--32.

\bibitem[Ferreira et~al.(2021)Ferreira, Pilastri, Martins, Pires, and
  Cortez]{Ferreira2021}
L.~Ferreira, A.~Pilastri, C.~M. Martins, P.~M. Pires, and P.~Cortez, ``{A
  Comparison of AutoML Tools for Machine Learning, Deep Learning and
  XGBoost},'' in \emph{2021 Int'l Joint Conf. on Neural Networks (IJCNN)}, vol.
  2021-July, no. October.\hskip 1em plus 0.5em minus 0.4em\relax IEEE, jul
  2021, pp. 1--8.

\bibitem[Balaji and Allen(2018)]{Balaji2018}
A.~Balaji and A.~Allen, ``{Benchmarking Automatic Machine Learning
  Frameworks},'' aug 2018.

\bibitem[Gunning et~al.(2019)Gunning, Stefik, Choi, Miller, Stumpf, and
  Yang]{Gunning2019xai}
D.~Gunning, M.~Stefik, J.~Choi, T.~Miller, S.~Stumpf, and G.-Z. Yang,
  ``Xai—explainable artificial intelligence,'' \emph{Science Robotics},
  vol.~4, no.~37, p. eaay7120, 2019.

\bibitem[Linardatos et~al.(2021)Linardatos, Papastefanopoulos, and
  Kotsiantis]{Linardatos2021}
P.~Linardatos, V.~Papastefanopoulos, and S.~Kotsiantis, ``{Explainable ai: A
  review of machine learning interpretability methods},'' \emph{Entropy},
  vol.~23, no.~1, pp. 1--45, 2021.

\bibitem[Novielli et~al.(2020)Novielli, Calefato, Dongiovanni, Girardi, and
  Lanubile]{Novielli2020}
N.~Novielli, F.~Calefato, D.~Dongiovanni, D.~Girardi, and F.~Lanubile, ``Can we
  use se-specific sentiment analysis tools in a cross-platform setting?'' in
  \emph{Proc. of 17th Int'l Conf. on Mining Software Repositories}.\hskip 1em
  plus 0.5em minus 0.4em\relax New York, NY, USA: ACM, 2020, p. 158–168.

\bibitem[Shaver et~al.(1987)Shaver, Schwartz, Kirson, and O'connor]{Shaver1987}
P.~Shaver, J.~Schwartz, D.~Kirson, and C.~O'connor, ``Emotion knowledge:
  further exploration of a prototype approach.'' \emph{Journal of personality
  and social psychology}, vol.~52, no.~6, p. 1061, 1987.

\bibitem[Ortu et~al.(2016)Ortu, Murgia, Destefanis, Tourani, Tonelli, Marchesi,
  and Adams]{Ortu2016}
M.~Ortu, A.~Murgia, G.~Destefanis, P.~Tourani, R.~Tonelli, M.~Marchesi, and
  B.~Adams, ``The emotional side of software developers in jira,'' in
  \emph{2016 IEEE/ACM 13th Working Conf. on Mining Software Repositories
  (MSR)}.\hskip 1em plus 0.5em minus 0.4em\relax IEEE, 2016, pp. 480--483.

\bibitem[Sebastiani(2002)]{Sebastiani2002}
F.~Sebastiani, ``Machine learning in automated text categorization,'' \emph{ACM
  Comput. Surv.}, vol.~34, no.~1, p. 1–47, mar 2002.

\bibitem[Calefato et~al.(2018)Calefato, Lanubile, Maiorano, and
  Novielli]{Calefato2018}
F.~Calefato, F.~Lanubile, F.~Maiorano, and N.~Novielli, ``Sentiment polarity
  detection for software development,'' \emph{Empirical Software Engineering},
  vol.~23, no.~3, pp. 1352--1382, 2018.

\bibitem[Zhang et~al.(2020)Zhang, Xu, Thung, Haryono, Lo, and Jiang]{Zhang2020}
T.~Zhang, B.~Xu, F.~Thung, S.~A. Haryono, D.~Lo, and L.~Jiang, ``Sentiment
  analysis for software engineering: How far can pre-trained transformer models
  go?'' in \emph{2020 IEEE Int'l Conf. on Software Maintenance and Evolution
  (ICSME)}.\hskip 1em plus 0.5em minus 0.4em\relax IEEE, 2020, pp. 70--80.

\bibitem[Crisan and Fiore{-}Gartland(2021)]{Crisan2021}
\BIBentryALTinterwordspacing
A.~Crisan and B.~Fiore{-}Gartland, ``Fits and starts: Enterprise use of automl
  and the role of humans in the loop,'' in \emph{{CHI} '21: {CHI} Conf. on
  Human Factors in Computing Systems, Virtual Event / Yokohama, Japan, May
  8-13, 2021}, Y.~Kitamura, A.~Quigley, K.~Isbister, T.~Igarashi, P.~Bj{\o}rn,
  and S.~M. Drucker, Eds.\hskip 1em plus 0.5em minus 0.4em\relax {ACM}, 2021,
  pp. 601:1--601:15. [Online]. Available:
  \url{https://doi.org/10.1145/3411764.3445775}
\BIBentrySTDinterwordspacing

\bibitem[Serban et~al.(2020)Serban, van~der Blom, Hoos, and Visser]{Serban2020}
\BIBentryALTinterwordspacing
A.~Serban, K.~van~der Blom, H.~Hoos, and J.~Visser. (2020) The 2020 state of
  engineering practices for machine learning. [Online]. Available:
  \url{https://se-ml.github.io/automl-report2020}
\BIBentrySTDinterwordspacing

\bibitem[Xin et~al.(2021)Xin, Wu, Lee, Salehi, and Parameswaran]{Xin2021}
D.~Xin, E.~Y. Wu, D.~J.-L. Lee, N.~Salehi, and A.~Parameswaran, ``{Whither
  AutoML? Understanding the Role of Automation in Machine Learning
  Workflows},'' in \emph{Proc. of 2021 CHI Conf. on Human Factors in Computing
  Systems}.\hskip 1em plus 0.5em minus 0.4em\relax ACM, may 2021, pp. 1--16.

\bibitem[Sun et~al.(2023)Sun, Song, Gui, Ma, and Wang]{Sun2023}
Y.~Sun, Q.~Song, X.~Gui, F.~Ma, and T.~Wang, ``{AutoML in The Wild: Obstacles,
  Workarounds, and Expectations},'' in \emph{Proc of 2023 CHI Conf. on Human
  Factors in Computing Systems (CHI '23), April 23{\^{a}}•ﬁ28, 2023,
  Hamburg, Germany}, vol.~1, no.~1.\hskip 1em plus 0.5em minus 0.4em\relax
  Association for Computing Machinery, feb 2023.

\bibitem[Pruitt and Grudin(2003)]{Pruitt2003personas}
J.~Pruitt and J.~Grudin, ``Personas: practice and theory,'' in
  \emph{Proceedings of the 2003 conference on Designing for user experiences},
  2003, pp. 1--15.

\end{thebibliography}

\end{document}